\documentstyle[prl,aps,epsf,multicol]{revtex}

\begin{document}

\draft
\preprint{{\bf ETH-TH/98-??}}

\title{Superfluidity versus Bloch oscillations in confined atomic gases}

\author{H.P.\ B\"uchler $^{a}$, V.B.\ Geshkenbein $^{ab}$, and G.\
  Blatter $^{a}$}

\address{$^{a}$Theoretische Physik, ETH-H\"onggerberg, CH-8093
  Z\"urich, Switzerland \\
$^{b}$Landau Institute for Theoretical Physics, 117940 Moscow, Russia.}

\date{\today}
\maketitle

\begin{abstract}
  {We study the superfluid properties of (quasi) one-dimensional
   bosonic atom gases/liquids in traps with finite geometries in the
   presence of strong quantum fluctuations. Driving the condensate
   with a moving defect we find the nucleation rate for phase
   slips using instanton techniques. While phase slips are quenched
   in a ring resulting in a superfluid response, they proliferate
   in a tube geometry where we find Bloch oscillations in the
   chemical potential. These Bloch oscillations describe the
   individual tunneling of atoms through the defect and thus are
   a consequence of particle quantization.}
\end{abstract}

\begin{multicols}{2}
\narrowtext

Bose-Einstein condensation\cite{Bose1924a} (BEC) and
superfluidity\cite{landau41} are basic characteristics of
Bosonic quantum gases and fluids. While the Bose-Einstein
condensate (of density $n_0$) is a thermodynamic quantity
characterizing off-diagonal-long-range {\it order}, the superfluid
density $n_s$ describes the {\it response} to a perturbation in
the broken phase \cite{penrose56}. In real quantum liquids, such
as bulk $^4$He, condensation and superfluidity appear in unison,
but in general one may be realized without the other. E.g.,
non-interacting Bose gases in three dimensions form a condensate
without superfluidity ($n_s = 0$), while confining a real quantum
fluid to two dimensions destroys the condensate ($n_0 = 0$) but
preserves superfluidity. In one dimension only superfluidity may
survive at zero temperature.

A major recent breakthrough is the realization of a Bose-Einstein
condensate in weakly interacting atomic gases. The confinement
within a trap has important consequences for the {\it condensate}
\cite{ketterle96,dalfovo99}: e.g., in a three dimensional harmonic
trap the well known ideal gas condensation temperature
$T_{\rm\scriptscriptstyle BE}=3.3\: \hbar^2 \nu^{2/3}/ m$ is
replaced by
$T_{\rm ho}^{\rm\scriptscriptstyle 3D} =0.94 \: \hbar
\omega_{\rm ho} N^{1/3}$.
Quite notable is the appearance of a condensate below
$T_{\rm ho}^{\rm\scriptscriptstyle 1D} = \hbar
\omega_{\parallel} N/\ln 2N$ in an ideal gas confined to a (quasi)
one-dimensional (1D) geometry (here, $\nu$, $N$ and $m$ are the bulk
density, particle number, and mass of the bosons, while
$\omega_{\rm ho}^{3}= \omega_{\parallel} \omega_{\perp}^{2}$ with
$\omega_{\perp}$ and $\omega_{\parallel}$ denoting the transverse
and longitudinal trapping frequencies). The appearance of such a
condensate at $T_{\rm ho}^{\rm\scriptscriptstyle 1D}$ is
characterized by a sharp crossover for an ideal 1D gas;
the broadening of the transition due to interparticle interactions
has been discussed by Petrov {\it et al.} \cite{petrov00}.

First attempts to probe the (bulk) {\it superfluid} properties of
condensed atom gases through a moving laser beam have been carried
out recently\cite{raman99}; the results on the critical velocity
are in rough agreement with expectations deriving from a weak
coupling analysis based on the Gross-Pitaevskii
theory\cite{frisch92}.  An interesting question then arises
regarding the interplay of superfluidity and enhanced
thermal/quantum fluctuations due to dimensional reduction. In this
letter we study the superfluid response of (quasi) one-dimensional
atomic gases and fluids trapped within finite tube and ring
geometries, see Fig.\ 1, where quantum phase slips tend to destroy
superfluidity. While interesting on their own, these questions
have attracted much attention recently through the novel atom chip
technology \cite{folman00} allowing for the experimental
realization of strongly confined atom gases exhibiting large
quantum fluctuations.
\vspace{-0.1cm}
\begin{figure}[htpp]
  \centerline{\epsfxsize= 8.0cm \epsfbox{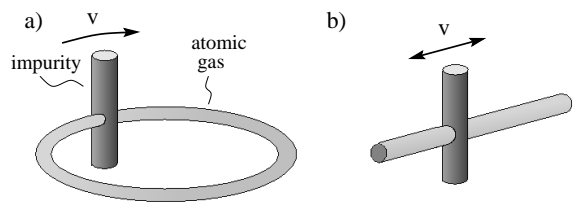}}
  \vspace{0.1cm}
  \caption{Trap geometries of the atomic gas perturbed by a
     moving laser beam: (a) ring structure with periodic
     boundary conditions, (b) finite length ($L$) tube with
     closed ends.}
  \label{superfluidring}
\end{figure}
\vspace{-0.3cm}
The destruction of dissipation free flow in one dimensional (1D)
superconductors and superfluids is triggered by the appearance of
quantum phase slips as discussed by Zaikin {\it et
al.}\cite{zaikin97} for metal wires and by Kagan {\it et
al.}\cite{kagan00} for super\-fluid rings.  Here, we model the
action of the laser beam through a moving impurity (velocity $v$)
and derive a low frequency effective action describing the
dynamics of the phase difference across. The quantum nucleation
rate for phase slips determines the response:
at finite temperature the infinite system exhibits a linear
response and hence is not superfluid, $\Delta \mu \propto v$
with $\Delta \mu$ the drop in the chemical potential across
the impurity.
This contrasts with the ring geometry where interactions quench
the phase slip nucleation below a critical velocity, establishing
a superfluid response.  In a finite tube the quantum phase 
slips proliferate and the new non-superfluid ground state 
exhibits Bloch oscillations in the chemical potential difference 
across the moving impurity, $\Delta\mu\propto\sin(2\pi n v t)$ 
with $n$ the 1D atom density. The physical origin of these 
oscillations is
found in the particle quantization: the moving impurity enhances
the particle density in front, producing a chemical potential
difference across the impurity, which in turn is released each
time an atom tunnels through the impurity.

Trapped 1D atomic gases have attracted much interest recently
\cite{petrov00,Olshanii_98,Kolomeisky_00,Girardeau_00}. Their
nature is conveniently described in terms of the dimensionless
parameter  $\gamma^{-1} = n\hbar^2/m g = n l_\perp^2/2a$.
Here, $n$ and $m$ denote the (one-dimensional) atom density
and mass, respectively, and
$l_\perp = \sqrt{\hbar/m \omega_\perp}$ is the transverse
extension of the ground state wave function; assuming a
contact potential with a scattering length $a < l_{\perp}$
the interaction acquires a 3D character and the interaction
parameter takes the form $g=2 \hbar^2 l_{\perp}^2/m a$, see
\cite{Olshanii_98} for corrections. Small and large values
of $\gamma$ correspond to weakly and strongly interacting
bosons: for weakly interacting gases the energy per particle
is given by the bosonic expression
$\epsilon_{\rm \scriptscriptstyle B}(n) = g n/2$
and the Gross-Pitaevskii equation holds. In the
strongly interacting situation ($\gamma \rightarrow \infty$)
the 1D Fermion-Boson duality becomes manifest and the
energy per particle is given by the fermionic expression
$\epsilon_{\rm\scriptscriptstyle F}(n) = \pi\hbar^2 n^{2}/6m$;
the implications for the density profile have
been discussed in \cite{Olshanii_98} and an appropriate
modification of the Gross-Pitaevskii equation has been proposed by
Kolomeisky {\it et al.} \cite{Kolomeisky_00}. It is the latter
limit describing impenetrable bosons with strong quantum
fluctuations we are mainly interested in the present work; the
requirements for the experimental realization of this so called
Tonks-Girardeau limit have been discussed in Ref.\
\cite{Olshanii_98}. A convenient starting point describing the low
energy physics in both cases is the imaginary time action for the phase
$\phi(x,\tau)$ of the bosonic field $\psi \propto \exp[ i\phi]$
\cite{lieb63.1},
\begin{equation}
  {\mathcal S}_{0} =\frac{K \hbar }{2
    \pi} \int_{0}^{\hbar \beta} d\tau \int
  dx  \left[c_{s} \left(\partial_{x} \phi \right)^{2} +\frac{1}{c_{s}}
    \left(\partial_{\tau} \phi \right)^{2} \right],
  \label{masslessfield}
\end{equation}
describing sound modes with velocity $c_{s}$ to be cut-off at high
energies $\Lambda$. For weak interaction, $\gamma \ll 1$, the
sound velocity $c_{s} =\sqrt{n g/m}$, the dimensionless parameter
$K = \pi/\sqrt{\gamma}$, and the cutoff $\Lambda = n g$ derive
from the Gross-Pitaevskii equation; quantum fluctuations are
suppressed in this limit. Increasing the interaction, quantum
fluctuations renormalize the sound velocity $c_{s}$ and the
dimensionless parameter $K$ in the low energy action
(\ref{masslessfield}): strongly interacting bosons with a contact
potential $g \delta(x)$ are described by $K \approx 1+4/\gamma$
while $c_s K = \pi n \hbar/m$ remains unrenormalized (note that
$K\rightarrow 1$ in the strong coupling or Tonks-Girardeau limit
$\gamma \rightarrow \infty$; parameters $K<1$ might be realized
for bosons with long range interactions \cite{lieb63.1}). In
Eq.~(\ref{masslessfield}) we assume a flat trapping potential
along the longitudinal direction. A weak longitudinal trapping
potential can be accounted for by space dependent parameters
$K(x)$ and $c_{s}(x)$; the resulting deformation of the excitation
spectrum will not change the main results presented below.
Equation (\ref{masslessfield}) produces the $T=0$ phase correlator
$\langle[\phi(x)-\phi(x')]^2\rangle \sim \ln|x-x'|/K$, hence phase
fluctuations destroy the condensate in the infinite system. On the
other hand, the logarithmic divergence of the phase correlator is
cut off in a finite trap, allowing for the definition of a
(quasi-) condensate even at finite (but low) temperatures as
discussed by Petrov {\it et al.} \cite{petrov00} (note that the
use of (\ref{masslessfield}) requires $T < \Lambda$).

The {\it superfluid} properties of the (quasi-) condensate can be
probed with a moving laser beam \cite{raman99} which we describe
by a (strong) impurity potential $V_{\scriptscriptstyle \rm
imp}(z,t) = g_{\scriptscriptstyle \rm imp} \delta(z-v t)$ with
$g_{\scriptscriptstyle \rm imp}> g$, suppressing the particle
density locally. In the weakly interacting limit we can integrate
the Gross-Pitaevskii equation over the impurity region and derive
a Josephson term coupling the left and right parts of the atom gas
(the `leads') \cite{giovanazzi00,us},
\begin{equation}
 V(\varphi) =  E_{J} \left[1
    -\cos\varphi(\tau) \right] - \hbar n v \varphi,
\end{equation}
with $E_{J}=(K/\pi)n g^{2}/g_{\scriptscriptstyle \rm imp}$, to be
renormalized in the presence of large interactions. The second
term drives the phase difference $\varphi$ across the impurity.
Small contributions from higher harmonics do not modify the
results below which are dominated by large scales; also, on the
level of (\ref{masslessfield}) such terms are irrelevant in the
renormalization group sense \cite{kane92}. The classical
(meta-)stable states $\varphi_{j}$ derive from minimizing
$V(\varphi)$ and fall into the intervals $\varphi_{j}-2\pi j \in
[-\pi,\pi)$; on a semi-classical level we define the associated
ground state $|j\rangle$ of the $j$-th well. Next, we integrate
over the phase dynamics (\ref{masslessfield}) in the leads
\cite{us} and obtain the effective action for the phase difference
$\varphi$ across the impurity
\begin{equation}
  {\mathcal S}= \hbar \int   \frac{d\omega}{8
    \pi^{2}} Q(\omega) |\varphi(\omega)|^{2}
 +  \int  d\tau  V\left[\varphi(\tau)\right].
  \label{effact}
\end{equation}
Extended leads produce an ohmic kernel $Q(\omega) \sim K|\omega|$
with a characteristic time scale $\tau_{c} \sim K \hbar /E_{J}$
for the phase dynamics; the finite leads in a trap geometry will
strongly modify the low frequency part of the kernel $Q$ with
dramatic consequences for the response.

Starting from the classical stationary states $\varphi_{j}$ describing
a superfluid system we have to account for quantum fluctuations introducing
transitions between these states which potentially destroy the superfluid
response. Indeed, depending on the low frequency dynamics encoded in
the behavior of the kernel $Q(\omega \rightarrow 0)$ the relevant (semi-classical)
instanton solutions \cite{coleman77} of (\ref{effact}) will provide us with
dissipative phase slips and a linear response in the infinite system,
coherent hopping and Bloch oscillations in the finite tube, or confinement
and hence superfluid response in the ring. A crucial element entering this
analysis is the nature of the quantum variable $\varphi$ itself: while
$\varphi \in {\bf R}$ is an extended variable if different minima $\varphi_j$
are physically distinguishable, $\varphi \in [0,2\pi)$ turns into a compact
variable if this is not the case. In the following we discuss the superfluid
response for the different geometries in more detail.

The {\it infinite} system is characterized by an ohmic kernel
$Q(\omega) = K |\omega|$; the effective action (\ref{effact}) describes a
particle in a periodic potential with damping $K/2 \pi$.
The tunneling between classical minima leads to the excitation of
sound modes rendering the states distinguishable; as a consequence
the phase $\varphi$ has to be treated as an extended variable
\cite{schoen90}. The system
exhibits a quantum phase transition at $K=1$ \cite{schmid83} separating
a non-superfluid ground state with a delocalized phase $\varphi$ at
$K < 1$ from a superfluid ground state with a localized phase at $K > 1$.
The finite temperature response at $K > 1$ is determined by the thermally
assisted quantum nucleation of phase slips \cite{weiss85}: the corresponding
action involves a kink-antikink--pair separated by the distance
$\overline{\tau}$ in imaginary time,
\begin{displaymath}
  \frac{{\mathcal S}}{\hbar} =
  K \ln\bigg[\Big( \frac{\hbar}{\pi T \tau_{c}}\Big)^{2}
    \sin^{2}\Big(\frac{\pi T \overline{\tau}}{\hbar}\Big) \bigg]
   - 2 \pi n v \overline{\tau} .
\end{displaymath}
Using instantons \cite{coleman77} the nucleation rate
is given by the integral $\Gamma \sim (1/\tau_{c}^{2})\int_{0}^{+i \infty}
d\overline{\tau} \exp(-{\mathcal S}/\hbar)$. We recover a superfluid
response at $T=0$ with an algebraic rate $\Gamma \sim v^{2K-1}$,
while the response turns linear at finite temperature, $\Gamma \sim v
T^{2K-2}$, as thermally activated quantum phase slips destroy the
phase coherence across the link.

The {\em ring} geometry (see Fig.~\ref{superfluidring}(a)) introduces
periodic boundary conditions for the phase, $\phi(x,\tau) = \phi(x+L,
\tau)$, with two important consequences: {\it i)} the existence of
a winding number defines an extended quantum variable $\varphi$,
and {\it ii)}, the sound modes are quantized and exhibit a self
interaction due to the compactness of the loop, modifying the kernel
at low frequencies,
\begin{equation}
 Q(\omega) = K  \omega \coth \frac{\omega L}{2 c_{s}} \simeq \left\{
    \begin{array}{l c}
      2 K c_{s}/L ,  \hspace{15pt}& \omega
     < 2 c_{s}/L, \\ & \vspace{-5pt}
     \\
    K |\omega| , & \omega > 2 c_{s}/L .
    \end{array}
      \right.
      \label{periodickernel}
\end{equation}
The static potential $(\hbar K c_{s}/\pi L) \varphi^{2}/2$
describes the kinetic energy of the flow in the ring and is easily
understood when the static solution $\phi(x) = \varphi\,
[1/2+x/L-\Theta(x)]$ is inserted in (\ref{masslessfield}). The
additional potential renders the system superfluid
\cite{hekking97}: the new minima satisfy the relation
$(E_{J}/\hbar n)\sin\varphi_{j}=v-v_{L}\varphi_{j}/\pi$, where the
first term is the usual flow induced by the motion of the
impurity, while the second term $\propto v_{L} = K c_{s}/nL$ is
due to the static potential. The absolute minimum at $\varphi_{j}$
with $|v-2 v_{L} j|< v_{L}$ describes a stable superfluid state
with a critical velocity $v_c = v_L$. Indeed, the $T=0$ action for
a kink-antikink--pair exhibits a linear confinement (we assume $K
> 1$),
\begin{displaymath}
  \frac{{\mathcal S}}{\hbar} =
  K \ln \bigg[ \Big(\frac{L}{\pi c_{s} \tau_{c}} \Big)^{2}
    \sinh \frac{\pi c_{s} \overline{\tau}}{L} \bigg]
- 2 \pi n (v\!-\!2 v_{L} j) \overline{\tau},
\end{displaymath}
and the nucleation of phase slips is quenched for $v < v_L$. At
finite temperatures or large drives $|v - 2v_{L} j| > v_{L}$
incoherent tunneling processes via thermally activated quantum
nucleation of phase slips describe the equilibration of the ring
towards its thermal equilibrium as given by the appropriate
density matrix. At very large drives $|v- 2v_{L} j| \gg v_{L}$ the
system is far from equilibrium and the response resembles that of
the infinite wire.

In a finite length {\em tube} the flow vanishes at the tube ends
providing us with the boundary conditions $\partial_{x}\phi(-L/2,
\tau) = \partial_{x} \phi(L/2,\tau)=0$. The finite size
quantization of the sound modes in the leads introduces a smooth
low frequency cutoff in the kernel,
\begin{equation}
Q(\omega) \!= \!K  \omega \tanh\frac{\omega L}{2 c_{s}}\!\simeq \! \left\{
    \begin{array}{l c} {\displaystyle
      K (L/2 c_{s}) \: \omega^{2}}  ,\!    & \omega
      < 2 c_{s}/L, \\ & \vspace{-5pt}
      \\
      K |\omega|, & \omega > 2 c_{s}/L ;
    \end{array}
  \right. \! \!\!
  \label{closedkernel}
\end{equation}
a simple understanding is provided by inserting the dynamic
solution $\phi(x,\tau) = \varphi(\tau)\,[1/2-\Theta(x)]$ into
(\ref{masslessfield}). The massive low frequency dynamics renders
the tunneling between the minima coherent and using instanton
techniques we can determine the hopping amplitude \cite{leggett87}
\begin{equation}
  W \sim  (\hbar/\tau_c) \left(c_s \tau_c/2 L\right)^K
  \label{W}
\end{equation}
between semi-classical states $|j\rangle$. In deriving (\ref{W})
we have assumed $L > c_{s} \tau_{c}$, see \cite{schoen90} for a
discussion of short systems. We have to distinguish between
`diagonal' transitions without excitations of sound modes and
`non-diagonal' ones changing the number of sound modes in the
leads: as long as the time evolution involves only diagonal
transitions the states $|j\rangle$ are indistinguishable and
$\varphi$ is a compact variable $\varphi \in [0,2\pi)$. However,
in the following analysis it is a matter of convenience to choose
$\varphi$ extended and compactify only at the end.

For $E_J/(hc_s/KL) > 1$ the amplitude $E_J$ is larger than the
plasma frequency of the well and we can study the system response
within a tight binding analysis. The action (\ref{effact}) can be
transcribed into the Hamiltonian
\begin{equation}
  H \! = \! - \frac{W}{2} \! \sum_{j} \! \big\{|j\rangle \langle j\!+\!1|
    \! +\! |j\!+\!1\rangle
    \langle j|\big\} \! -\! \epsilon
  \sum_{j} j |j\rangle \langle j|,
  \label{tbhamiltion}
\end{equation}
where the last term $\propto \epsilon= 2 \pi \hbar n v$ describes
the driving force ($\varphi \equiv 2 \pi \sum_{j} j |j\rangle \langle j|$
in the site basis $|j\rangle$). Applying the unitary transformation $U=
\exp\left[-2 \pi \: i \: {\mathcal N}(t) \varphi \right]$ with
${\mathcal N}(t)= n v t + {\mathcal N}_{0}$ we eliminate the drive
through a redefinition of the hopping amplitudes,
\begin{equation}
  H(t) = - \frac{W}{2} \sum_{j} \left\{e^{ 2 \pi \: i \: {\mathcal N}(t)}
|j\rangle \langle j+1| + {\rm c.c.}\right\}.  \label{gtbhamiltion}
\end{equation}
This Hamiltonian is equivalent to that of an electron in a crystal
driven by an electric force $e E\cong\hbar n v$ described by the
vector potential $e A/c=e E t \cong n v t$: at zero drive the
energy eigenstates form a Bloch band, while a finite
electric field leads to `Bloch oscillations'\cite{krieger86} (the
above transformation corresponds to a gauge transformation, and
the `quasi-number' ${\mathcal N}_{0}$ accounts for the gauge
freedom). At low temperatures and low drives no sound
waves are excited in the leads and we identify  $|0\rangle =
|j\rangle$ (the compact character of the phase restricts its value
to a region centered around the potential minimum at
$\varphi=0$ and tends to establish phase coherence; on the other
hand, the quantum nature of the phase and the presence of phase
slips give a finite probability to any value $\varphi \in [0,
2\pi)$, thus reducing the phase coherence across the impurity,
see also \cite{pitaevskii01}).
The Hamiltonian (\ref{gtbhamiltion}) reduces to $H=-W \cos\left[2
\pi {\mathcal N}(t) \right] |0\rangle \langle 0|$ and admits the
solution $|0\rangle(t) = \exp\left[ - i \int dt H(t)\right]
|0\rangle$. The state of the system and its energy depend on the
impurity position via the `quasi-number' ${\mathcal N}$; the usual
derivatives provide us with the chemical potential difference
across the impurity
\begin{equation}
\Delta \mu = \langle \partial_{{\mathcal N}} H\rangle
= 2 \pi W \sin \big[2 \pi {\mathcal N}(t)\big]
\label{dmu}
\end{equation}
and its time evolution due to the drive $v$
\begin{equation}
\partial_t {\mathcal N}  = n v.
\label{dtN}
\end{equation}
In the static limit with $v = 0$ the spectrum maps out a Bloch
band $E({\mathcal N}) = - W \cos(2\pi{\mathcal N})$
(the `quasi-number' ${\mathcal N}$ plays the role of the
`quasi-momentum' $k$ in a periodic crystal), while a finite
driving force $n v$ leads to `Bloch oscillations' in the
chemical potential difference $\Delta\mu=2\pi W\sin(2\pi n v t)$.
These oscillations are due to the accumulation of particles in front
of the impurity, the latter allowing only discrete particles to
tunnel.  Each `Um\-klapp' process describes a particle tunneling
through the impurity. The behavior of the tube then is dual to that
of the classic Josephson junction, as is easily seen when replacing
${\mathcal N}$ by the phase drop $\Phi$ and $\Delta \mu$ by
the supercurrent $I$: the relations (\ref{dmu}) and (\ref{dtN})
are equivalent to Josephson's famous relations $I = I_{c}\sin\Phi$
and $\partial_{t} \Phi = 2eV/\hbar$ with $I_c$ and $V$ the
critical current and voltage across the junction.

For high temperatures and drives processes involving frequencies
larger than $c_{s}/L$ induce nondiagonal transitions which compete
with the diagonal ones. `Bloch oscillations' then disappear above
the crossover temperature $T_{L} = \hbar c_{s}/L$ and the critical
drive $v_{L}/K$. At high temperatures $T \gg T_{L}$ or high drives
$v\gg v_{L}/K$ all processes are fast and we recover the physics
of the infinite wire with incoherent tunneling via the quantum
nucleation of phase slips (we assume $K>1$).

The quantum nucleation of phase slips leads to a transfer of
energy to the bosonic system at high drives $v>v_{L}$
but well below the mean field critical velocity $E_{J}/\hbar n$.
Then the {\it macroscopic quantum tunneling} of the phase can be
observed via the heating of the sample, in analogy to the
experiment by Raman {\it et al.} \cite{raman99} (note that our
work predicts a dissipation free low-drive response and
the appearance of a critical velocity for both topologies,
ring and tube). On the other hand, the Bloch oscillations at
low drives constitute a {\it macroscopic quantum coherence}
phenomenon leading to density fluctuations within the leads.
Using a second laser beam to probe the
oscillating densities in the leads allows to measure these
fluctuations, at least in principle. However, as each `Umklapp'
process involves only one particle tunneling through the impurity
these oscillations will be small, thus requiring a high sensitivity
in the experiment.

In conclusion, geometric confinement of the atom gas boosts the
importance of fluctuations. The superfluid
response strongly depends on the particular geometry: in a ring
the phase difference across an impurity is well defined and the
response remains superfluid below the critical velocity
$v_{L} \propto 1/L$,
while in a tube phase slips proliferate and driving
the system induces `Bloch oscillations' in the chemical potential
across the impurity.

We thank A.I.\ Larkin, W.\ Zwerger, and W.\ Ketterle for stimulating discussions.

\end{multicols}

\vspace{-0.2cm}
\begin{thebibliography}{10}
\vspace{-1.2cm}

\bibitem{Bose1924a}
S.\ Bose,
\newblock Z.\ Phys. {\bf 26}, 178 (1924);
\newblock A.\ Einstein,
\newblock Sitzungber.\ Preuss.\ Akad.\ Wiss.\ {\bf 1925}, 3 (1925).

\bibitem{landau41}
L.D.\ Landau,
\newblock J.\ Phys.\ U.S.S.R.\ {\bf 5}, 71 (1941).

\bibitem{penrose56}
O.\ Penrose and L.\ Onsager,
\newblock Phys.\ Rev. {\bf 104}, 576 (1956).

\bibitem{ketterle96}
W.\ Ketterle and N.J.\ van Druten,
\newblock Phys.\ Rev.\ A {\bf 54}, 656 (1996).

\bibitem{dalfovo99}
F.\ Dalfovo, S.\ Giorgini, L.P.\ Pitaevksii, and S.\ Stringari,
\newblock Rev.\ Mod.\ Phys. {\bf 71}, 463 (1999).

\bibitem{petrov00}
D.S.\ Petrov, G.V.\ Shlyapnikov, and J.T.M.\ Walraven,
\newblock Phys.\ Rev.\ Lett. {\bf 85}, 3745 (2000).

\bibitem{raman99}
C.\ Raman et~al.,
\newblock Phys.\ Rev.\ Lett. {\bf 83}, 2502 (1999).

\bibitem{frisch92}
T.\ Frisch, Y.\ Pomeau, and S.\ Rica,
\newblock Phys.\ Rev.\ Lett. {\bf 69}, 1644 (1992);
\newblock B.\ Jackson, J.F.\ McCann, and C.S.\ Adams,
\newblock Phys.\ Rev.\ A {\bf 61}, 051603 (2000);
\newblock J.S.\ Stiessberger and W.\ Zwerger,
\newblock Phys.\ Rev.\ A {\bf 62}, 061601 (2000).

\bibitem{folman00}
R.\ Folman et~al.,
\newblock Phys.\ Rev.\ Lett. {\bf 84}, 4749 (2000);
\newblock J.\ Reichel, W.\ H{\"a}nsel, and T.W.\ H{\"a}nsch,
\newblock Phys.\ Rev.\ Lett. {\bf 83}, 3398 (1999).

\bibitem{zaikin97}
A.D.\ Zaikin, D.S.\ Golubev, A.\ van Otterlo, and G.T.\ Zim\'anyi,
\newblock Phys.\ Rev.\ Lett. {\bf 78}, 1552 (1997).

\bibitem{kagan00}
Y.\ Kagan, V.N.\ Prokof'ev, and B.V.\ Svistunov,
\newblock Phys.\ Rev.\ A {\bf 61}, 045601 (2000);
see also V.A.\ Kashurnikov et al.,
Phys.\ Rev.\ B {\bf 53}, 13091 (1996).

\bibitem{Olshanii_98} M.\ Olshanii,
   Phys.\ Rev.\ Lett.\ {\bf 81}, 938 (1998), see also V.\ Dunjko
   {\it et al.}, cond-mat/0103085.

\bibitem{Kolomeisky_00} E.B.\ Kolomeisky {\it et al.},
   Phys.\ Rev.\ Lett.\ {\bf 85}, 1146 (2000).

\bibitem{Girardeau_00} M.D.\ Girardeau and E.M.\ Wright,
   Phys.\ Rev.\ Lett.\ {\bf 84}, 5239 (2000).

\bibitem{lieb63.1}
E.H.\ Lieb and W.\ Liniger,
\newblock Phys.\ Rev.\ {\bf 130}, 1605 (1963);
\newblock K.B.\ Efetov and A.I.\ Larkin,
\newblock Sov.\ Phys.\ JETP {\bf 42}, 390 (1975);
\newblock F.D.M.\ Haldane,
\newblock Phys.\ Rev.\ Lett.\ {\bf 47}, 1840 (1981);
\newblock V.N.\ Popov,
\newblock {\em Functional Integrals in Quantum Field Theory and Statstical
  Physics},
\newblock D.\ Reidel Publishing Company, 1983.


\bibitem{giovanazzi00}
S.\ Giovanazzi, A.\ Smerzi, and S.\ Fantoni,
\newblock Phys.\ Rev.\ Lett. {\bf 84}, 4521 (2000).

\bibitem{us} H.P.\ B\"uchler, V.B.\ Geshkenbein, and
G.\ Blatter (unpublished).

\bibitem{kane92}
C.L.\ Kane and M.P.A.\ Fisher,
\newblock Phys.\ Rev.\ Lett.\ {\bf 68}, 1220 (1992).

\bibitem{coleman77}
S.\ Coleman,
\newblock Phys.\ Rev.\ D {\bf 16}, 2929 (1977).

\bibitem{schoen90}
G.\ Sch{\"o}n and A.D.\ Zaikin,
\newblock Phys.\ Rep.\ {\bf 198}, 237 (1990).

\bibitem{schmid83}
A.\ Schmid,
\newblock Phys.\ Rev.\ Lett.\ {\bf 51}, 1506 (1983).

\bibitem{weiss85}
U.\ Weiss and H.\ Grabert,
\newblock Phys.\ Lett.\ {\bf 108A}, 63 (1985).

\bibitem{hekking97}
F.W.J.\ Hekking and L.I.\ Glazman,
\newblock Phys.\ Rev.\ B {\bf 55}, 6551 (1997).

\bibitem{leggett87}
A.J.\ Leggett et~al.,
\newblock Rev.\ Mod.\ Phys.\ {\bf 59}, 1 (1987).

\bibitem{krieger86}
J.\ Krieger and G.\ Iafrate,
\newblock Phys.\ Rev.\ B {\bf 33}, 5494 (1986).

\bibitem{pitaevskii01}
L.\ Pitaevskii, and S.\ Stringari,
\newblock cond-mat/0104458.

\end{thebibliography}
\end{document}